\documentclass[epj,nopacs]{svjour}
\usepackage{graphics}
\usepackage{graphicx}
\usepackage{color}
\usepackage{latexsym}
\usepackage[colorlinks,linkcolor=blue,citecolor=blue,urlcolor=black]{hyperref}

\newcommand{\be}{\begin{eqnarray}}
\newcommand{\ee}{\end{eqnarray}}

\newcommand{\cb}{\textcolor{black}}

\begin{document}
\title{Iron line reverberation mapping in Ghasemi-Nodehi-Bambi background}
\author{M. Ghasemi-Nodehi\inst{1}  
\thanks{\emph{email:} mghasemin@ipm.ir}%
}                     
%
%
\institute{School of Astronomy, Institute for Research in Fundamental Sciences (IPM), P. O. Box 19395-5531, Tehran, Iran}
\date{Received: date / Revised version: date}
%
\abstract{
The reverberation associated with the iron line is the time lag between direct photons from the corona and the photons reflected from the disk. The resulting line spectrum is called the 2D transfer function. The shape of the 2D transfer function is determined by the geometry of spacetime and the properties of BH. In a paper (Ghasemi-Nodehi and Bambi, EPJC 76 (2016) 290), the authers have proposed a parametrization. This parametrization is aimed to test the Kerr nature of astrophysical black hole candidates. In this paper, I provide a reverberation mapping of the Ghasemi-Nodehi-Bambi metric in order to constrain the parameter of spacetime. 
All parameters can be constrained with the exception of $b_{11}$. The parameter $b_4$ is harder to constrain too. 
\PACS{
      {PACS-key}{discribing text of that key}   \and
      {PACS-key}{discribing text of that key}
     } 
} 
\maketitle
%

\section{Introduction}

Einstein proposed the theory of General Relativity (GR) over a century ago \cite{einstein1916}. While largely successful in the weak-field tests~\cite{will,will2,will3,will4}, the validity of GR in the strong field regime is still unexplored. For instance, the motions of pulsars are sensitive to strong-gravitational self-field effects~\cite{kr} 
According to the predictions of GR the formation of astrophysical BHs from gravitational collapse is well understood by the Kerr spacetime paradigm. The Kerr spacetime is a stationary, axisymmetric, asymptotically flat solution of vacuum Einstein equations. It is only determined by the mass and spin parameters \cite{kerr1,kerr2}. Because of the presence of a highly ionized environment the charge reaches equilibrium and its value is negligible. 

There are different scenarios in which the metric around astrophysical black holes (BHs) may deviate from the Kerr case. The problem is that there is a degeneracy between the parameters of Kerr spacetime and theories deviating from the Kerr case. In order to verify the Kerr solution of GR, it is not only enough to detect the Kerr properties by observations but also any deviations from the Kerr case should be ruled out. Recently, there have been a lot of efforts to study confirmation of GR and constraining the deviation from GR using both electromagnetic radiation and gravitational waves \cite{r1,r2,r3,r4,r5}. The recent detection of gravitational waves from the coalescence of BH binaries can explore the dynamical strong field regime, but the available data do not constrain significantly the possible deviation from GR~\cite{gw1,gw2,gw3}. The constraining power of gravitational waves has been shown in~\cite{gw4}.

In order to write a more general metric than the Kerr metric, one can parametrize the Kerr solution and then try to constrain the deviation from the Kerr case. There are several parametrized metric with their own advantages and disadvantages \cite{r2,r3,r5,p1,p2,p3,p4,p5,p6,p7,p8,p10,p11,gbma,p14}. 

In Ref. \cite{gbma} the authors proposed a parametrized metric in which one can recover the Kerr case by setting the deformation parameters equal to one. We already studied BH shadow and X-ray reflection spectroscopy of this metric in \cite{gbma} and \cite{gbiron}, respectively. The study of quasi-periodic oscillation QPO observations of this metric is in progress. In the present paper, we study the reverberation mapping of the metric. 

In the framework of the corona-disk model, reverberation is associated with the iron line is the time lag between direct photons from the  corona and the photons reflected from the disk. The resulting line spectrum is called the 2D transfer function. The shape of the 2D transfer function is determined by the geometry of spacetime and the properties of a BH. Thus, an accurate measurement of the 2D transfer function can be used as a probe of the spacetime geometry ~\cite{revf1,revf2,crev1,crev2}.

In this paper, first, I have studied the impact of the deformation parameters $b_i$ on the shape of the 2D transfer function. Then, using a minimum $\chi^2$ approach, I study the contour levels to constrain the parameters. Our reference BH is a Kerr BH and total photon count is $10^3$, which is for current observations. Except for $b_4$ and $b_{11}$, the parameters of the Ghasemi-Nodehi-Bambi (GB) metric can be constrained. $b_4$ and $b_{11}$ introduce a parameter degeneracy. However, higher spin values can remove the degeneracy of parameter $b_4$. I also checked the contours for the reference being a non-Kerr GB BH. The results are similar to the cases with the reference Kerr BH.
I also checked if it is possible to constrain the height of the corona considering that I have spin parameter from an independent measurement. As is shown, the height of the corona can be constrained except for the case $b_{11}$. 

I also have studied the BH shadow of the GB metric \cite{gbma} and X-ray reflection spectroscopy of the metric \cite{gbiron}. In shadow studies, the parameters $b_2, b_8, b_9$ and $b_{10}$ leave a small signature in the shadow boundary and $b_4, b_5, b_7$ and $b_{11}$ do not produce any specific signature in the shadow shape. In iron line studies, in the presence of the correct astrophysical model, 200 ks observations with future observational facilities such  as LAD/eXTP can constrain all the Kerr parameters except for $b_{11}$. The impact of $b_{11}$ on the iron line profile is extremely weak. I also try to provide possible constraints on the parameters using QPO, which is still in progress.

The content of the paper as follows. Section~\ref{rev} is devoted to the iron line reverberation mapping. The geometry of spacetime is discussed in section~\ref{gb}. Section~\ref{chi} shows our $\chi^2$ calculations in this paper.
Section~\ref{res} is for the results and discussions. Summary and conclusions are in section~\ref{summary}.


\section{Iron line reverberation mapping \label{rev}}

I use the accretion disk-corona model; the disk is on the equatorial plain orthogonal to the BH spin. 
The accretion disk emits as blackbody locally and as multicolor blackbody when integrated radially. The so-called corona is a hotter, usually optically thin electron cloud enshrouding the accretion disk. Its exact geometry is unknown, also in this regard some work has been done~\cite{geoco}. The thermal photons from the accretion disk can interact with hot electrons in the corona. Because of inverse Compton scattering the corona becomes an X-ray source with a power-law spectrum. The corona works as a point source located on the axis of the accretion disk just above the BH. This arrangement is known as the lamppost geometry of the disk-corona model~\cite{dc1,dc2}. However, a different geometry can be considered~\cite{diffco1,diffco2}. Here I consider the simple lamppost geometry. The disk-corona model is described by parameter the $h$ as the height of the corona above the disk in addition to the parameters of the BH spacetime. Furthermore, the inner edge is at ISCO. 
A photon of the corona enters the disk and may produce fluorescence emission line also referred to as the reflection component. 
The strongest line is the iron $K\alpha$ line at $\sim 6.4$ keV. Here I only consider the iron line. The coronal flux received by the disk obeys a power-law, as $r^{-q}$, where r is the disk radius; here I consider $q = 3$, which recovers the Newtonian limit at large distances but it might be different at small radii, $r \approx h$.

Here I consider reverberation associated with the iron line. It is the time lag between direct photons from the corona and the photons reflected from the disk. The resulting line spectrum, which is a function of both time and photon energy, is called the 2D transfer function. The shape of the 2D transfer function is different for different geometries and BH spacetimes. In addition to the fundamental properties of BH, it also depends on the height of the corona above the disk, $h$, and the inclination angle of the disk with respect to the observer's line of sight, $i$.

\cb{The time delay or lag is caused by the difference in light travel time between primary emission and reprocessed emission.
I also calculate the frequency dependence and the energy dependence of the lag. First, I calculate the response function for different $b_i$.  To plot the response, it is assumed that the rest frame spectrum is simply a $\delta$-function iron line at energy $6.4$ keV. Then, in order to calculate the frequency dependence of the lag, I follow the approach considered in~\cite{Cack}. First, I take the Fourier transform of the transfer function. The transfer function in the frequency domain is
\be
\Psi (f) = \int_0^{\infty} \psi (\tau)\,\, e^{-i 2 \pi f \tau} d\tau
\ee
where $\psi (\tau)$ is the transfer function in the time domain. The phase difference, $\phi$, is as follows~\cite{Cack}:
\be
\phi (f) = \textrm{tan}^{-1} \left( \frac{\textrm{Im}(\Psi)}{1+\textrm{Re}(\Psi)} \right).
\ee
The time lag is $\phi / 2 \pi f$. For the energy dependence of the lag I calculate the response of the disk at every energy. To obtain the energy dependence of the lag one should consider a frequency range to plot. The results are discussed in section~\ref{res}. 
}

\section{Ghasemi-Nodehi-Bambi spacetime}\label{gb}

In Ref.~\cite{gbma} one proposed a new parametrization to the Kerr metric. As another parametrization we want to constrain possible deviations from the Kerr solution of GR. Here we recover the Kerr case when all deformation parameters are equal to 1. Meanwhile, in other metrics the deformation parameters are additive and reduce to the Kerr case for vanishing deformation parameters. We want to see how mass and spin in metric component deform the spacetime. We introduce 11 new parameters in front of any mass and/or spin term. Any deviation from 1 deforms spacetime more or less from that of the prediction of GR. The metric is as follows:

\be\label{eq-m}
ds^2 &=& - \left( 1 - \frac{2 b_1 M r}{r^2 + b_2 a^2 \cos^2\theta} \right) dt^2 
\nonumber\\ &&
- \frac{4 b_3 M a r \sin^2\theta}{r^2 + b_4 a^2 \cos^2\theta} dt d\phi 
+ \frac{r^2 + b_5 a^2 \cos^2\theta}{r^2 - 2 b_6 M r + b_7 a^2} dr^2 
\nonumber\\ &&
+ \left( r^2 + b_8 a^2 \cos^2\theta \right) d\theta^2 
\nonumber\\ &&
+ \left( r^2 + b_9 a^2 + \frac{2 b_{10} M a^2 r 
\sin^2\theta}{r^2 + b_{11} a^2 \cos^2\theta} \right) \sin^2\theta d\phi^2 \, .
\ee
The metric reduces to the Kerr metric for the $b_i = 1$ for all $i$. Here we set $b_1 = b_3 = b_6 = 1$. $b_1$ is equal to 1 because it is the coefficient of the mass and $b_3 = 1$ in the same way;  $b_3 a$ is the asymptotic specific angular momentum. $b_6$ is close to 1 from solar system experiments. We do not consider these three parameters in our reverberation mapping calculations.

The primary aim of the parametrization to this metric was to see how each part of the metric provides a signature on observations. 
Moreover, Ref.~\cite{sch} provides an example of the usage of this metric. One considered a special class of quintessential Kerr black holes. This class generated a modification to Kerr geometry that apparently is an extension of the modification of the Kerr geometry represented by Ghasemi-Nodehi-Bambi (GB). This modification is because of a special class of quintessential fields. One discusses shadow and spectral line of this modification to the Kerr geometry.


\section{$\chi^2$ calculation for comparison of Kerr and GB background}\label{chi}

Here I follow the approach of \cite{crev1} for comparison of the reverberation transfer functions of the Kerr and non-Kerr backgrounds. I first consider a primary model with spin $a_*$, parameter $b_i$, viewing angle $i$, emissivity profile $q$, and height of the corona $h$. For the 2D transfer function I use the notation \cite{crev1}
\be
n_{jk} = n(a_*, b_i,i,q,h)
\ee
for the photon flux number density in the energy bin $[E_j, E_j + \Delta E]$  and in the time bin $[t_k, t_k + \Delta t_k]$ . The secondary model with parameters $a_*' , b_i' , i' , q',  h'$($n_{jk}' = n(a_*', b_i',i',q',h')$) would be compared with the primary model by introducing a normalized negative log-likelihood \cite{crev1},
\be \label{log-l1}
\mathcal{L} &=& \frac{1}{\sum_{j,k} n_{jk}} \left[
\sum_{j,k} \frac{\left(n_{jk} - \alpha n_{jk}'\right)^2}{n_{jk}}
\right] \, , \label{log-l2}
\ee 
where 
\be
\alpha = \frac{\sum_{j,k} n_{jk}'}{\sum_{j,k} n_{jk}'^2/n_{jk}} \, .
\ee

The corresponding $\chi^2$ is $N \mathcal{L}$. N is the number of detected photons. Here I consider $N = 10^3$, which is for high quality observation today. Furthermore, the $\Delta E$ here is $50$ eV and $\Delta t = M$. If I consider $M = 10^6 M_{\odot}$, $\Delta t$ would be about $5$s.

I simulate the 2D transfer function with $N$ photons.  The simulation is an extension of the code described in~\cite{code1,code2}. I also added Poisson noise to my data. I treat my simulations as mocked data and apply the $\chi^2 \sim N \mathcal{L}$ approach to comparing the data. The contours show $1-\sigma, 2-\sigma$ and $3-\sigma$ levels.


\section{Results and discussion}\label{res}

\begin{figure*}
\vspace{0.4cm}
\begin{center}
\includegraphics[type=pdf,ext=.pdf,read=.pdf,width=8.0cm]{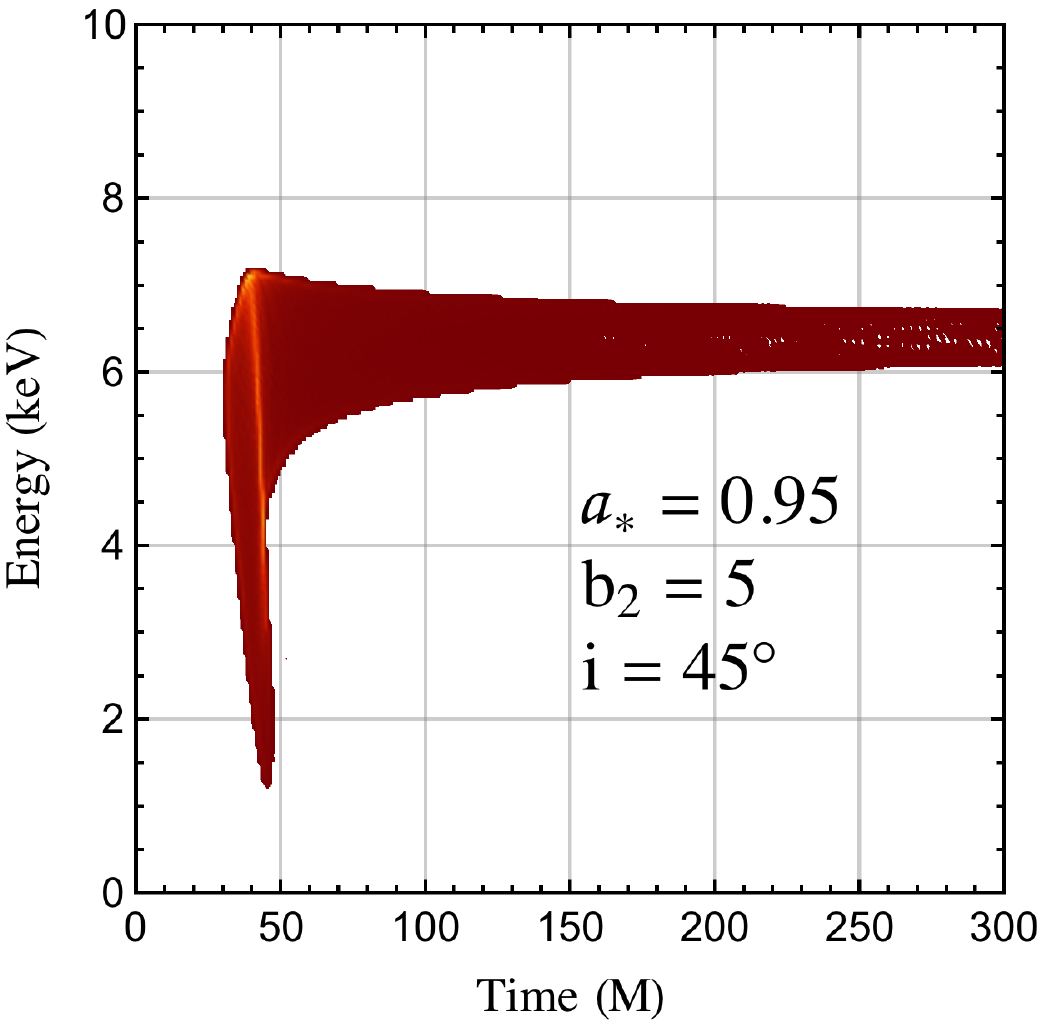}
\hspace{0.8cm}
\includegraphics[type=pdf,ext=.pdf,read=.pdf,width=8.0cm]{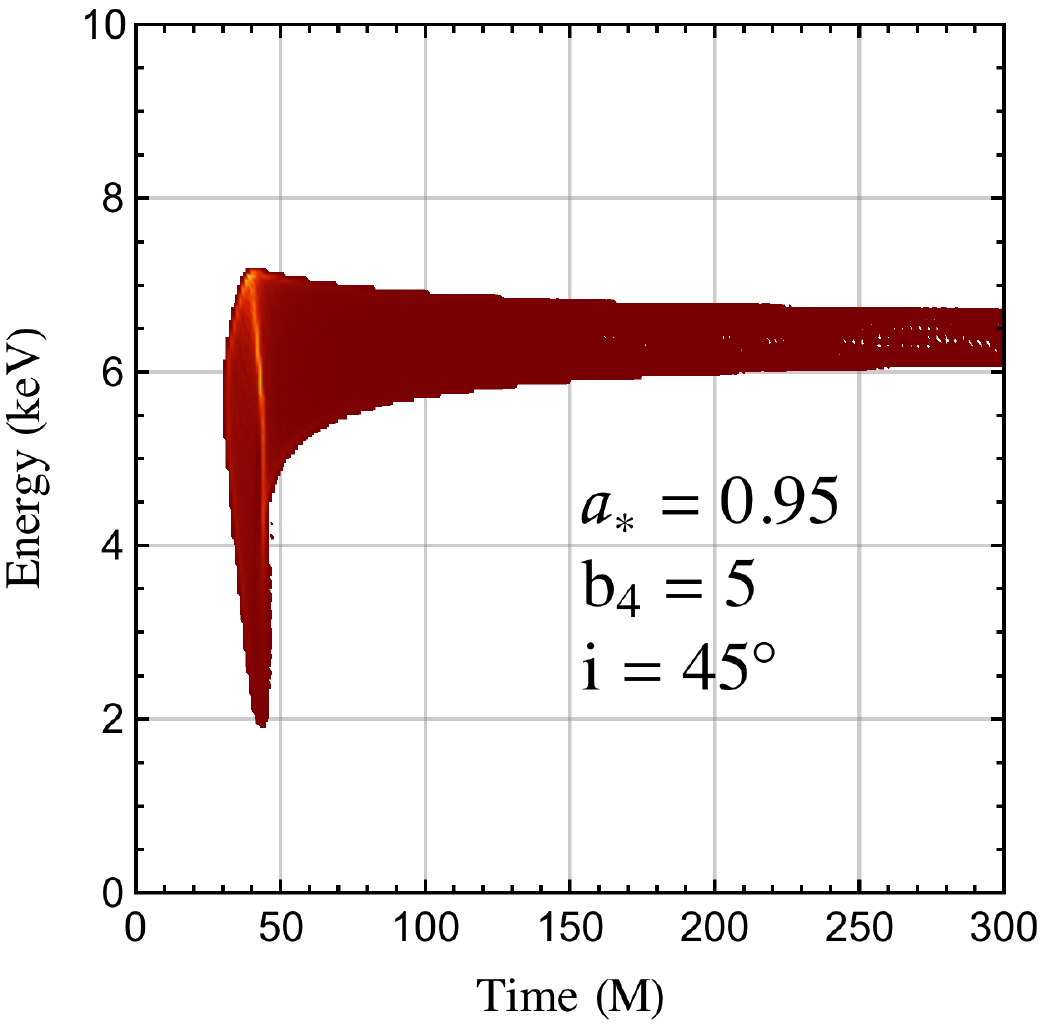}\\
\vspace{0.8cm}
\includegraphics[type=pdf,ext=.pdf,read=.pdf,width=8.0cm]{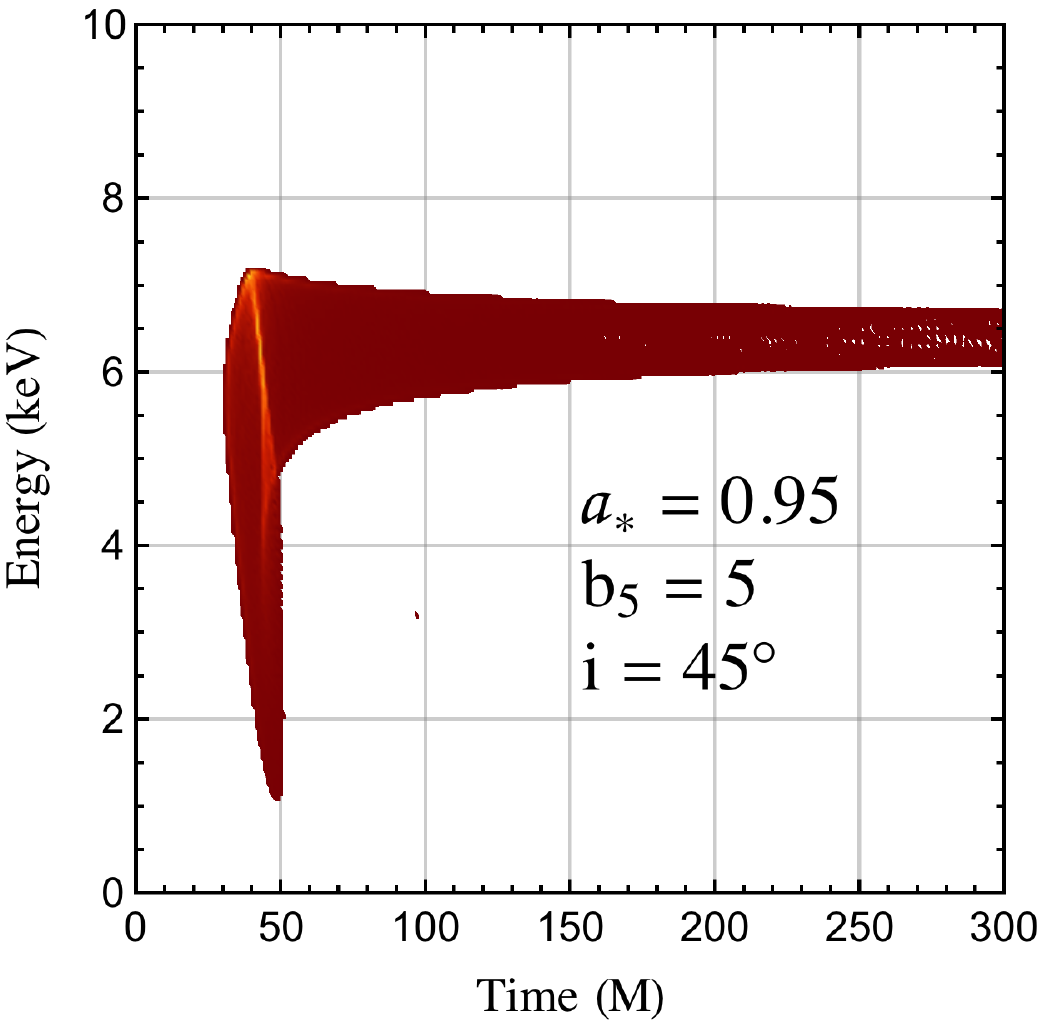}
\hspace{0.8cm}
\includegraphics[type=pdf,ext=.pdf,read=.pdf,width=8.0cm]{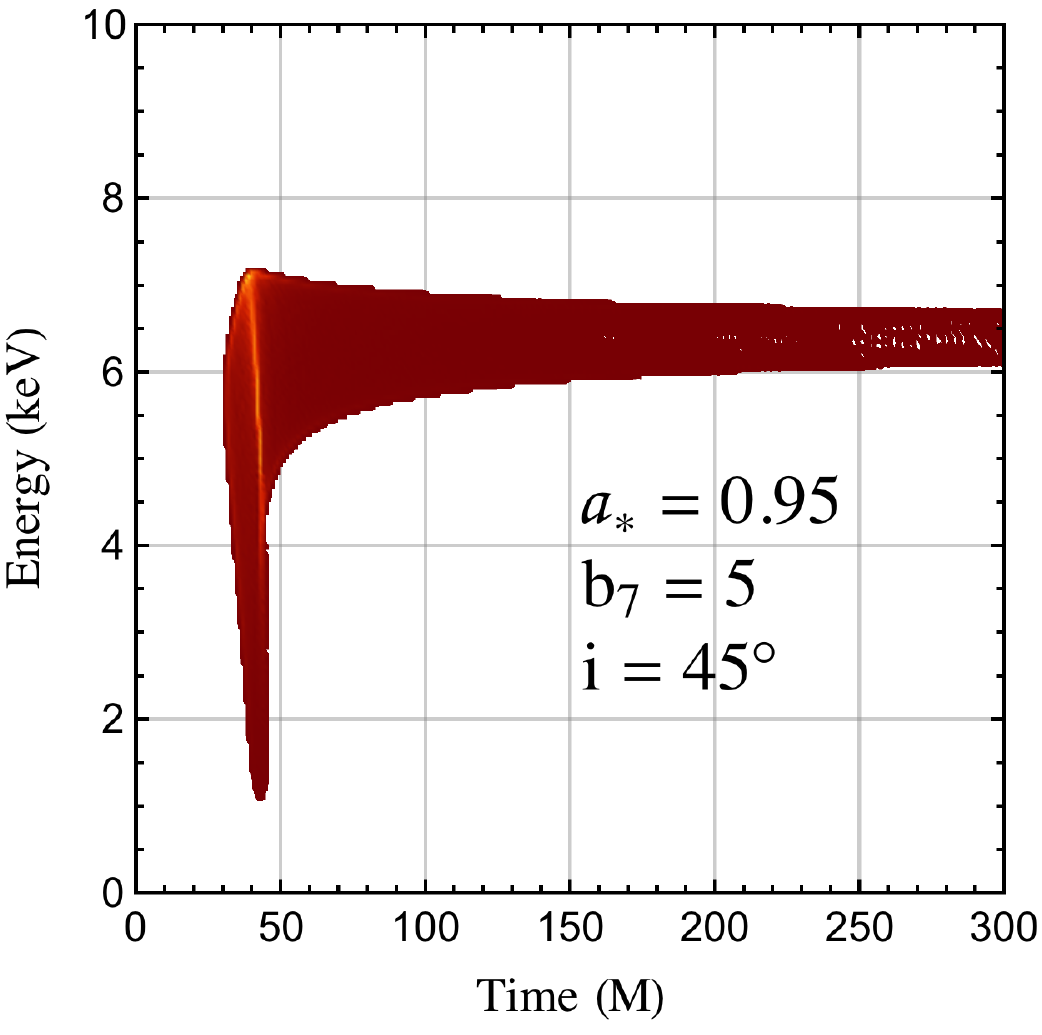}
\end{center}
\vspace{-0.3cm}
\caption{Reverberation mapping measurement associated with iron line in GB spacetime. Impact of the parameters $b_2$ (top left panel), $b_4$ (top right panel), $b_5$ (bottom left panel), $b_7$ (bottom right panel) on the 2D transfer function. See the text for more details. \label{rev1}}
\end{figure*}

\begin{figure*}
\vspace{0.4cm}
\begin{center}
\includegraphics[type=pdf,ext=.pdf,read=.pdf,width=8.0cm]{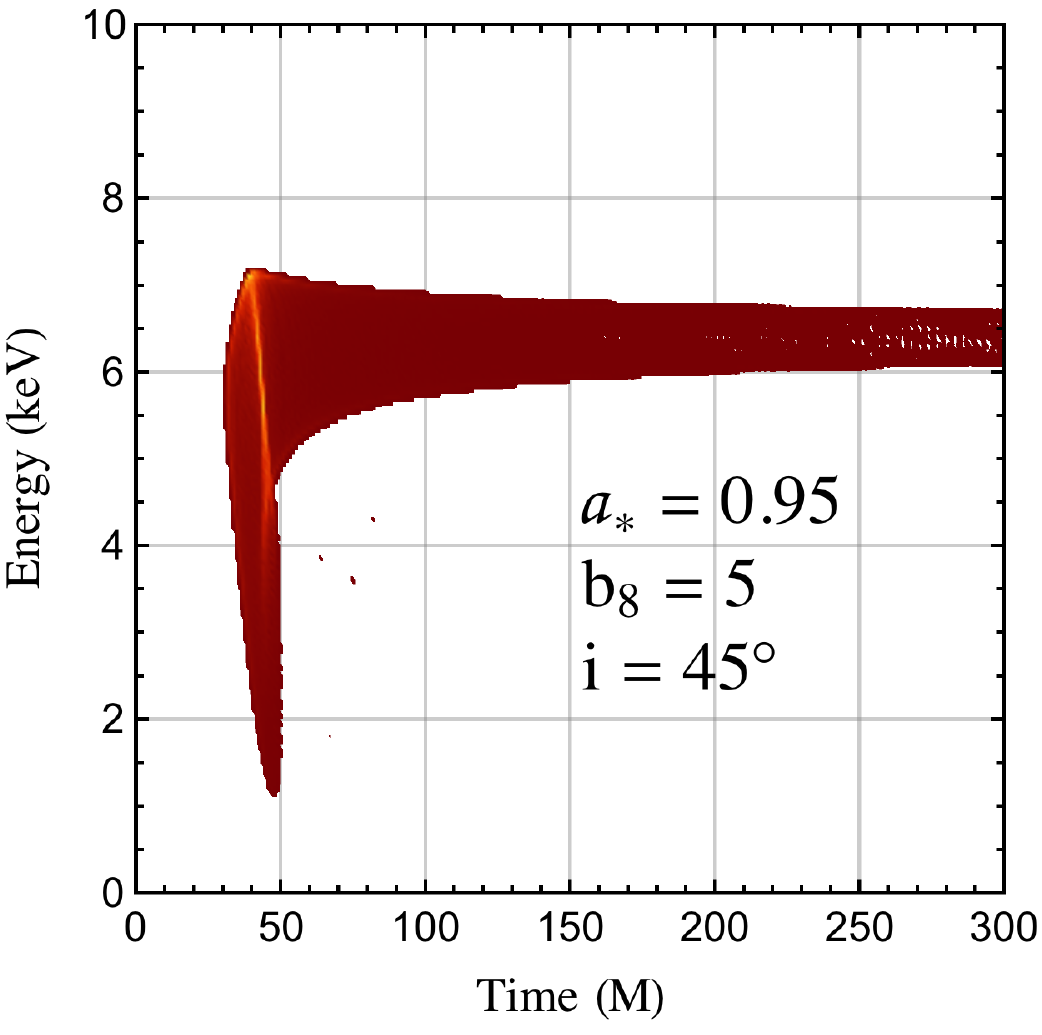}
\hspace{0.8cm}
\includegraphics[type=pdf,ext=.pdf,read=.pdf,width=8.0cm]{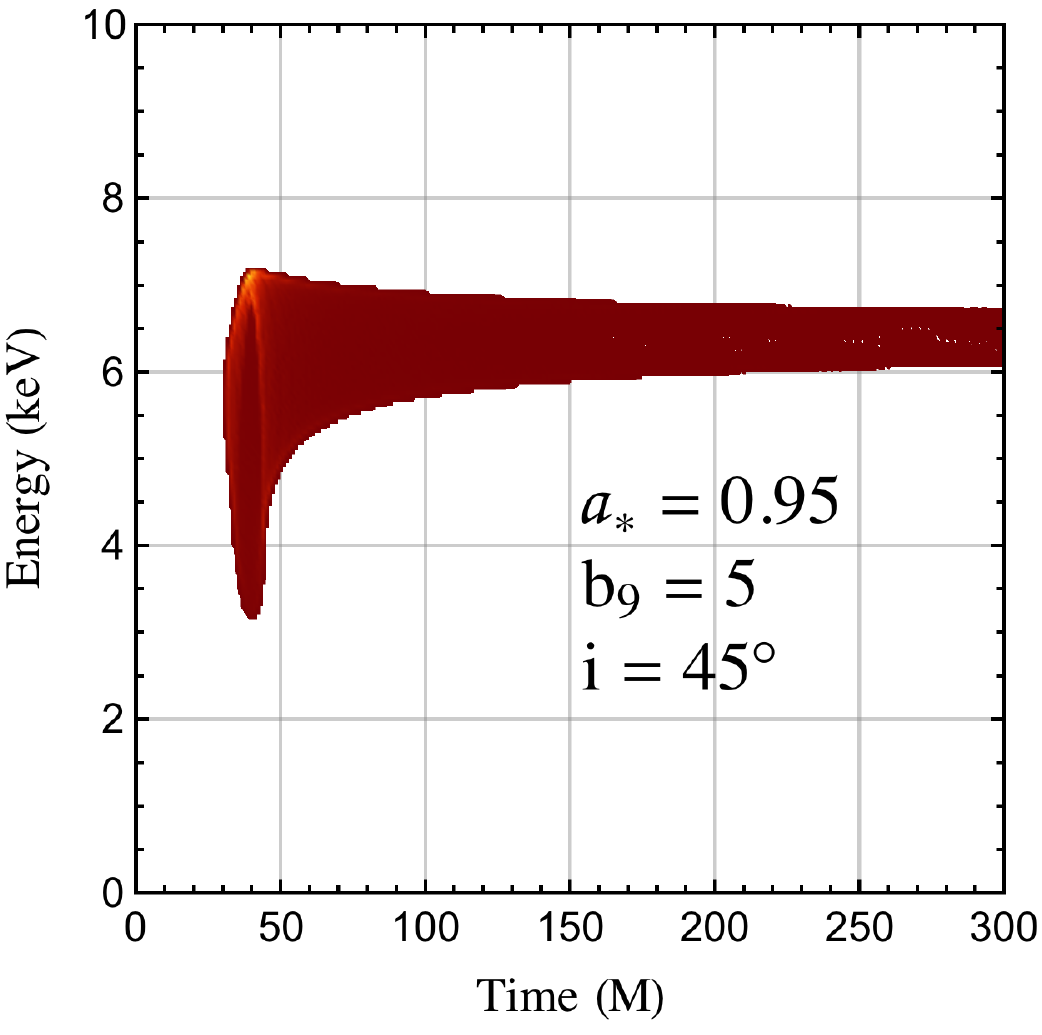}\\
\vspace{0.8cm}
\includegraphics[type=pdf,ext=.pdf,read=.pdf,width=8.0cm]{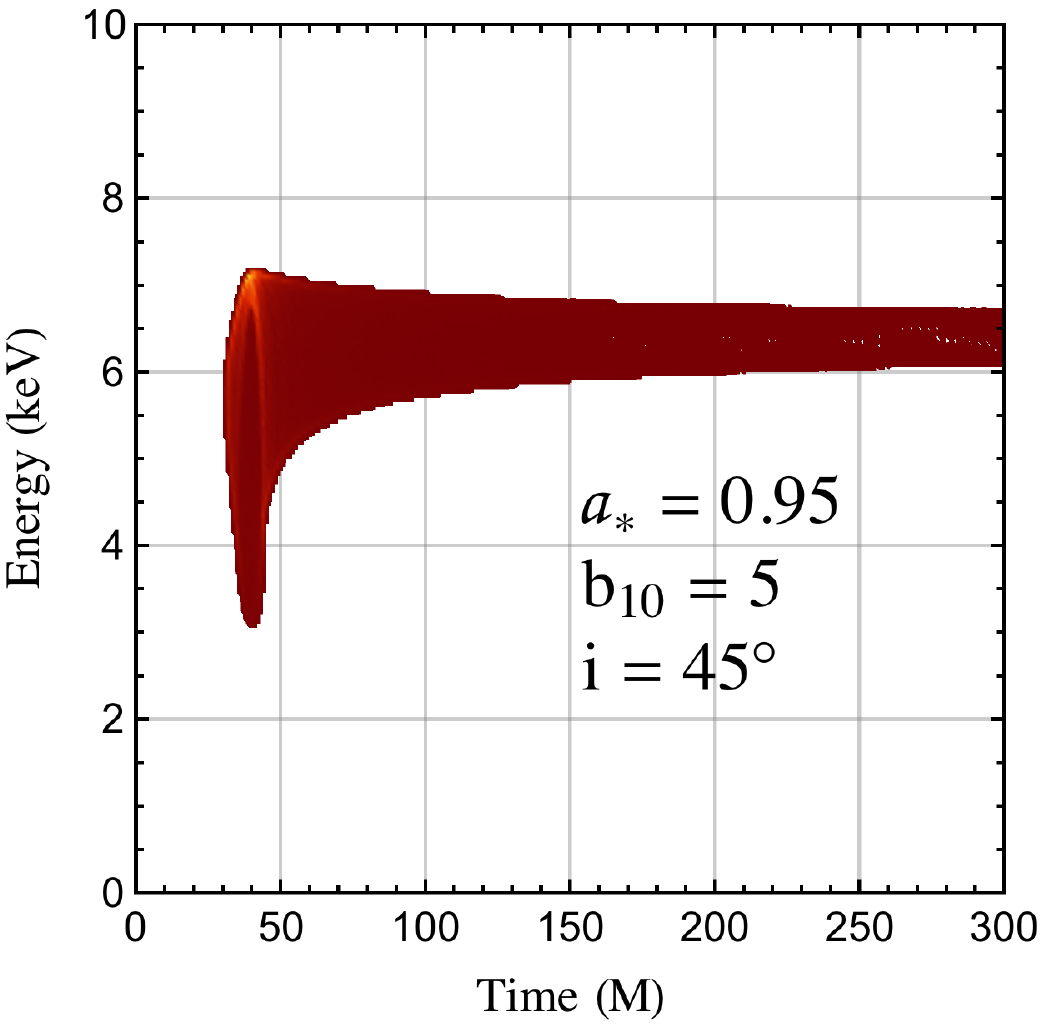}
\hspace{0.8cm}
\includegraphics[type=pdf,ext=.pdf,read=.pdf,width=8.0cm]{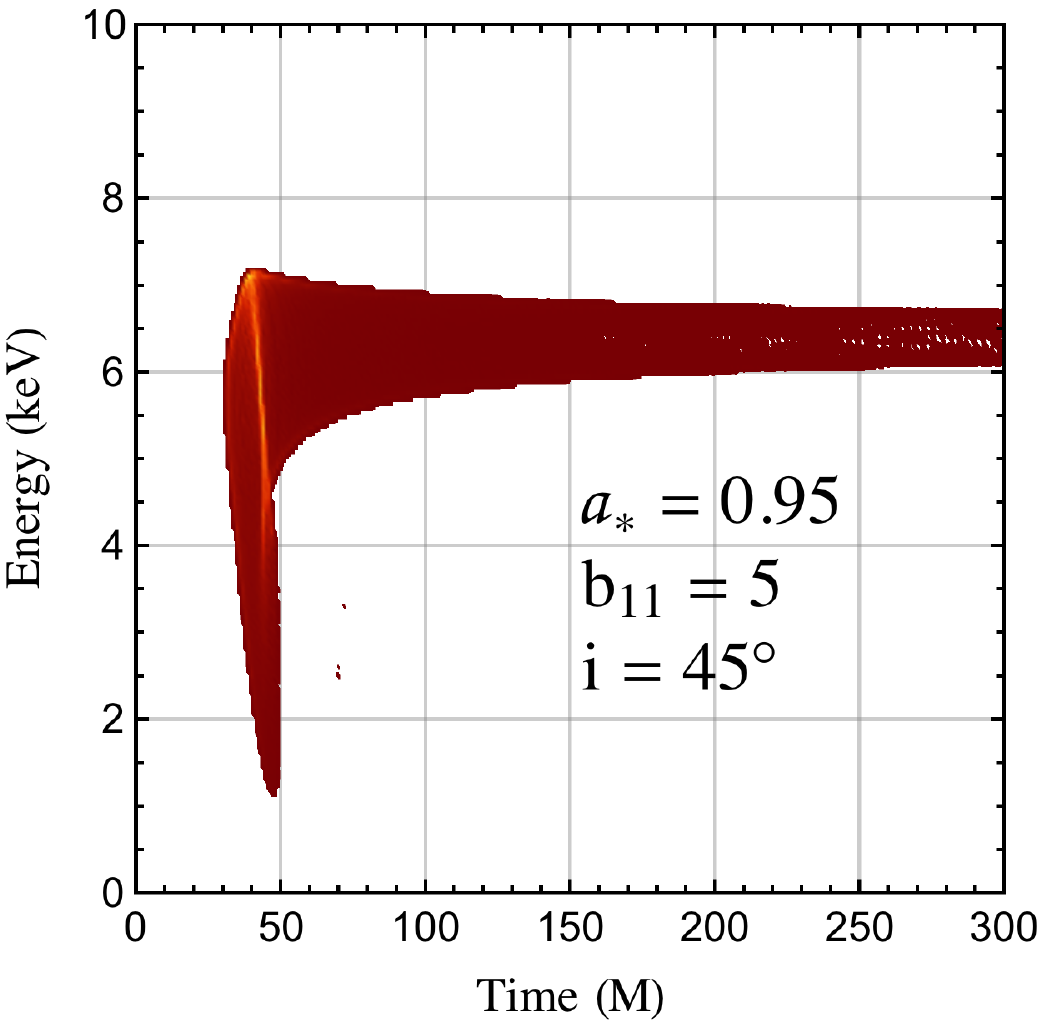}
\end{center}
\vspace{-0.3cm}
\caption{Reverberation mapping measurement associated with iron line in GB spacetime. Impact of the parameters $b_8$ (top left panel), $b_9$ (top right panel), $b_{10}$ (bottom left panel), $b_{11}$ (bottom right panel) on the 2D transfer function. See the text for more details. \label{rev2}}
\end{figure*}

\begin{figure*}
\vspace{0.4cm}
\begin{center}
\includegraphics[type=pdf,ext=.pdf,read=.pdf,width=10.0cm]{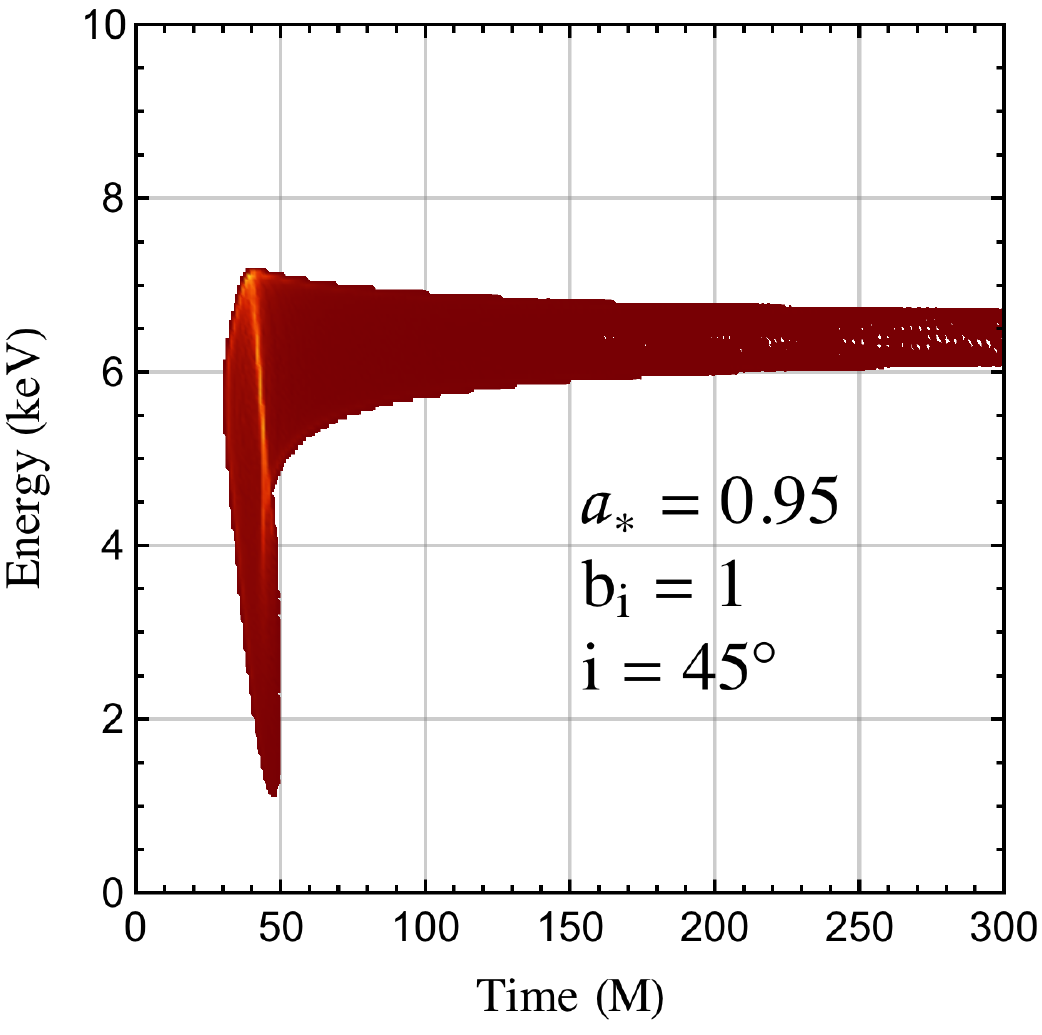}
\end{center}
\vspace{-0.3cm}
\caption{Reverberation mapping measurement associated to iron line in Kerr spacetime. See the text for more details. \label{revkerr}}
\end{figure*}

\begin{figure*}
\vspace{0.4cm}
\begin{center}
\includegraphics[type=pdf,ext=.pdf,read=.pdf,width=10.0cm]{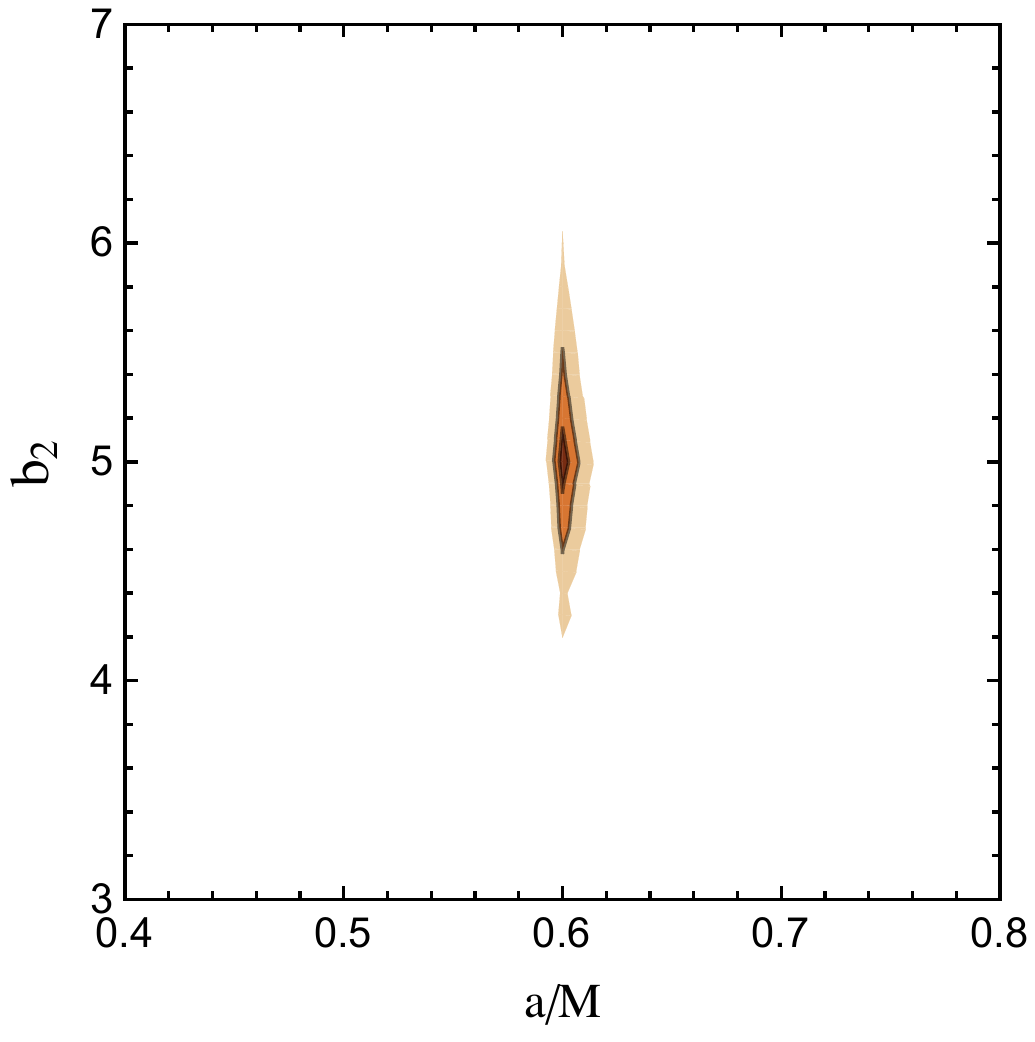}
\end{center}
\vspace{-0.3cm}
\caption{ Contour level from analysis of the 2D transfer function with GB BH as reference BH. Contour level is for parameters $b_2 = 5$, all other $b_i$s are equal to one. The reference model is a GB BH with $a_*' = 0.6$ and $i' = 55^{\circ}$. See the text for more details..\label{b2}}
\end{figure*}

\begin{figure*}
\vspace{0.4cm}
\begin{center}
\includegraphics[type=pdf,ext=.pdf,read=.pdf,width=10.0cm]{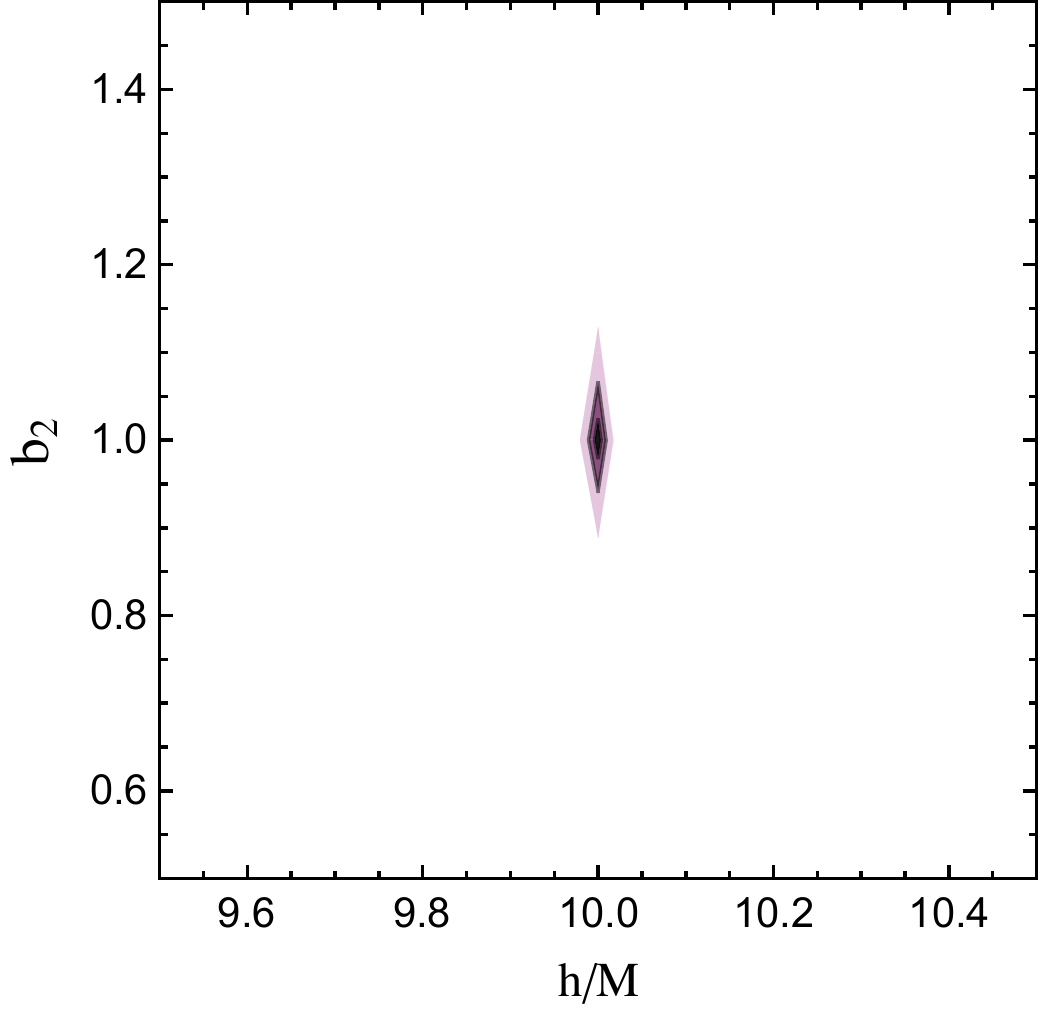}
\end{center}
\vspace{-0.3cm}
\caption{$b_2$ vs h contour levels from analysis of the 2D transfer function. The reference model is a Kerr BH with $a_*' = 0.9, h' = 10$ and $i' = 55^{\circ}$. See the text for more details. \label{he1}}
\end{figure*}

The results of my simulation of the 2D transfer function are shown in Figs~\ref{rev1}-\ref{he1}. In Figs.~\ref{rev1} and~\ref{rev2} , I have drawn the impact of the deformation parameter $b_i$ on the 2D transfer function.  I set h as $10$ and the viewing angle as $45^{\circ}$. Also the spin parameter is $0.95$ for all cases. I consider one $b_i = 5$ in each plot of Figs.~\ref{rev1} and~\ref{rev2} and all other $b_i = 1$. In this plot of the 2D transfer function the color indicates the photon number density. Fig.~\ref{revkerr} shows the 2D transfer function in Kerr space time. The spin parameter is $0.95$ and the viewing angle is $45^{\circ}$. 

I study contour levels as discussed in section~\ref{chi}. In the first study, the reference model is a Kerr black hole with $a_*' = 0.6, h' = 10, q' = 3$ and $i' = 55^{\circ}$. In all cases the total photon count is $N = 10^3$. According to contour studies the reverberation mapping can constrain $b_i$ except for $b_4$ and $b_{11}$ . The parameters $b_4$ and $b_{11}$ are degenerate. This degeneracy means that Kerr results can be reproduced by the parameters $b_4$ and $b_{11}$ of the GB metric. But in the case of $b_4$ the degeneracy would be removed by considering a higher spin value, $0.9$, but the case for $b_{11}$ is still degenerate even for a high spin value and also for a non-Kerr reference model. 
The values for the $b_i$ are as follows: 
\be
b_2 &=& 1^{+0.089}_{-0.049} ,\, \,
b_5 = 1^{+0.055}_{-0.089} \nonumber \\
b_7 &=& 1^{+0.348}_{-0.029} ,\, \,
b_8 = 1^{+0.086}_{-0.111} \nonumber \\
b_9 &=& 1^{+0.033}_{-0.041} , \, \,
b_{10} = 1^{+0.043}_{-0.067} \nonumber \\
\ee

Second, I also study the contours with the GB reference model for $b_i \neq 1$. The reference BH is a GB BH with spin $0.6$, the viewing angle $55^{\circ}$ and $b_i = 5$ for each case. Similar to the case with the Kerr BH as a reference, except for $b_4$ and $b_{11}$, all other parameters can be constrained. 
The value for $b_i = 5$ is as follows:
\be
b_2 &=& 5^{+0.070}_{-0.055} ,\, \,
b_5 = 5^{+0.764}_{-0.093} \nonumber \\
b_7 &=& 5^{+0.033}_{-0.041} ,\, \,
b_8 = 5^{+0.107}_{-0.113} \nonumber \\
b_9 &=& 5^{+0.025}_{-0.066} , \, \,
b_{10} = 5^{+0.193}_{-0.0001} \nonumber \\
\ee

Next, I consider a spin value of $0.9$ from independent observation. The reference height of the corona is set to 10. I try to see if I can constrain the height of the corona. The height of the corona can be constrained in all cases, with the exception of $b_{11}$. The case for $b_{11}$ seems more challenging. 
The value for the height of the corona is as follows:
\be
h = 10^{+0.001}_{-0.002}
\ee

As an example, the plot for the contour level from the analysis of the 2D transfer function with GB BH as the reference BH with $b_2 = 5$,  $a_*' = 0.6$ and $i' = 55^{\circ}$ is drawn in Fig.~\ref{b2}. Also, $b_2$ vs h contour levels from the analysis of the 2D transfer is shown in Fig.~\ref{he1}. The reference model is a Kerr BH with $a_*' = 0.9, h' = 10$ and $i' = 55^{\circ}$

As we see reverberation mapping can constrain the $b_i$ parameters except for $b_{11}$ and also the $b_4$ is harder to constrain. The power of reverberation measurement for constraining deviations  from GR is clear.

\begin{figure*}
\vspace{0.4cm}
\begin{center}
\includegraphics[type=pdf,ext=.pdf,read=.pdf,width=12.0cm]{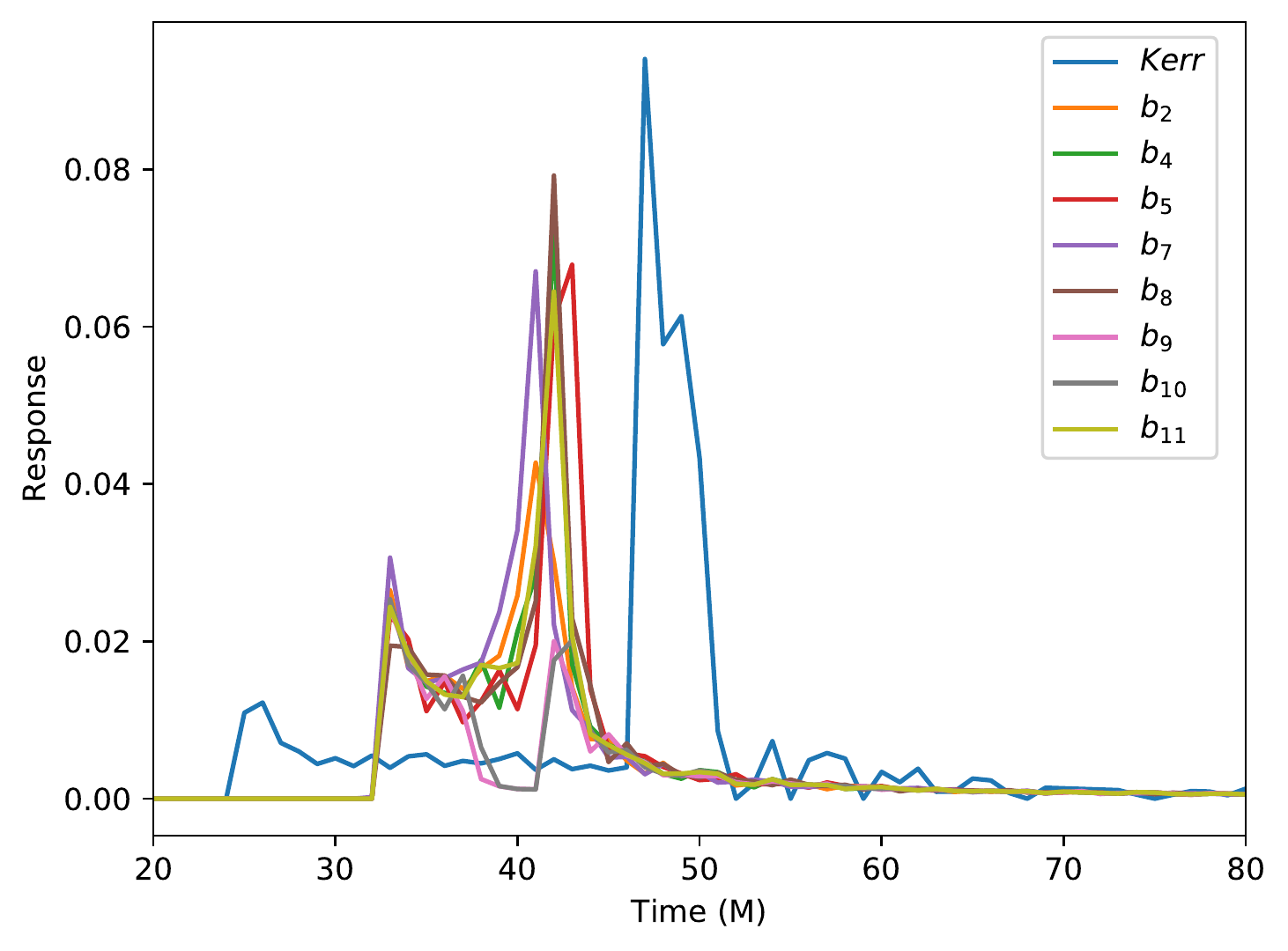}
\end{center}
\vspace{-0.3cm}
\caption{Response function for different $b_i = 5$. The spin parameter is $0.95$ and viewing angle is $45^{\circ}$. The height of the corona is $10$. \label{resps}}
\end{figure*}

\cb{In Fig.~\ref{resps}, I plot response function for different $b_i$. We see the responses are not similar to the Kerr case, so we expect these parameters can be constrained. From the figure, the response shifts to earlier time. We will better see this difference in the frequency and energy dependence of the lag. }

\begin{figure*}
\vspace{0.4cm}
\begin{center}
\includegraphics[type=pdf,ext=.pdf,read=.pdf,width=12.0cm]{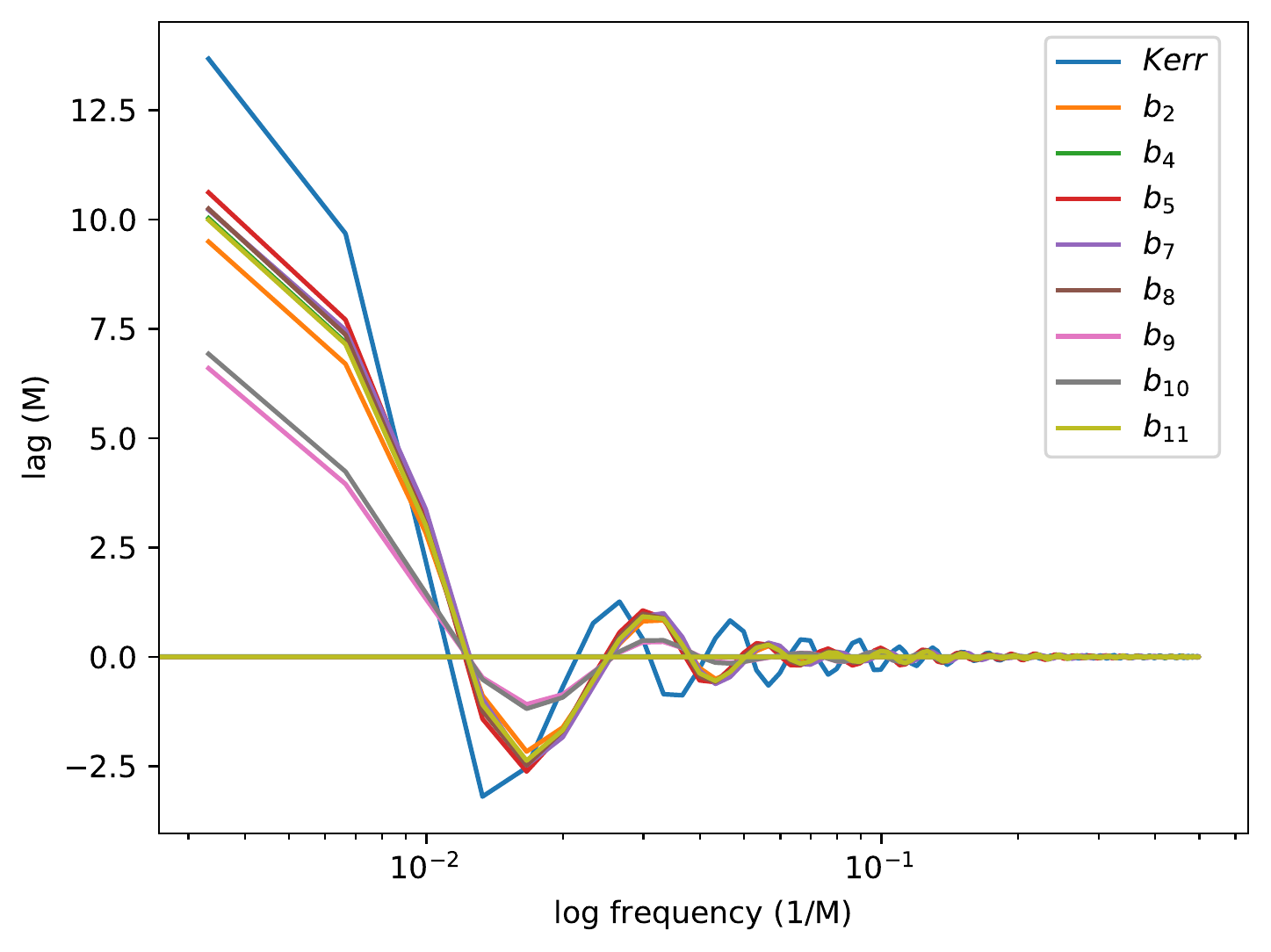}
\end{center}
\vspace{-0.3cm}
\caption{Lag vs frequency for different $b_i = 5$. The spin parameter is $0.95$ and viewing angle is $45^{\circ}$. The height of the corona is $10$. \label{freq}}
\end{figure*}

\cb{I have also drawn the lag vs frequency in Fig.~\ref{freq}. As we see the lag starts from a value and then oscillates around zero. There is significant difference from the Kerr case but the cases for $b_2, b_4, b_5, b_7, b_8, b_{11}$ are very similar. The cases for $b_9$ and $b_{10}$ have less lag at low frequencies than the other parameters. }

\begin{figure*}
\vspace{0.4cm}
\begin{center}
\includegraphics[type=pdf,ext=.pdf,read=.pdf,width=12.0cm]{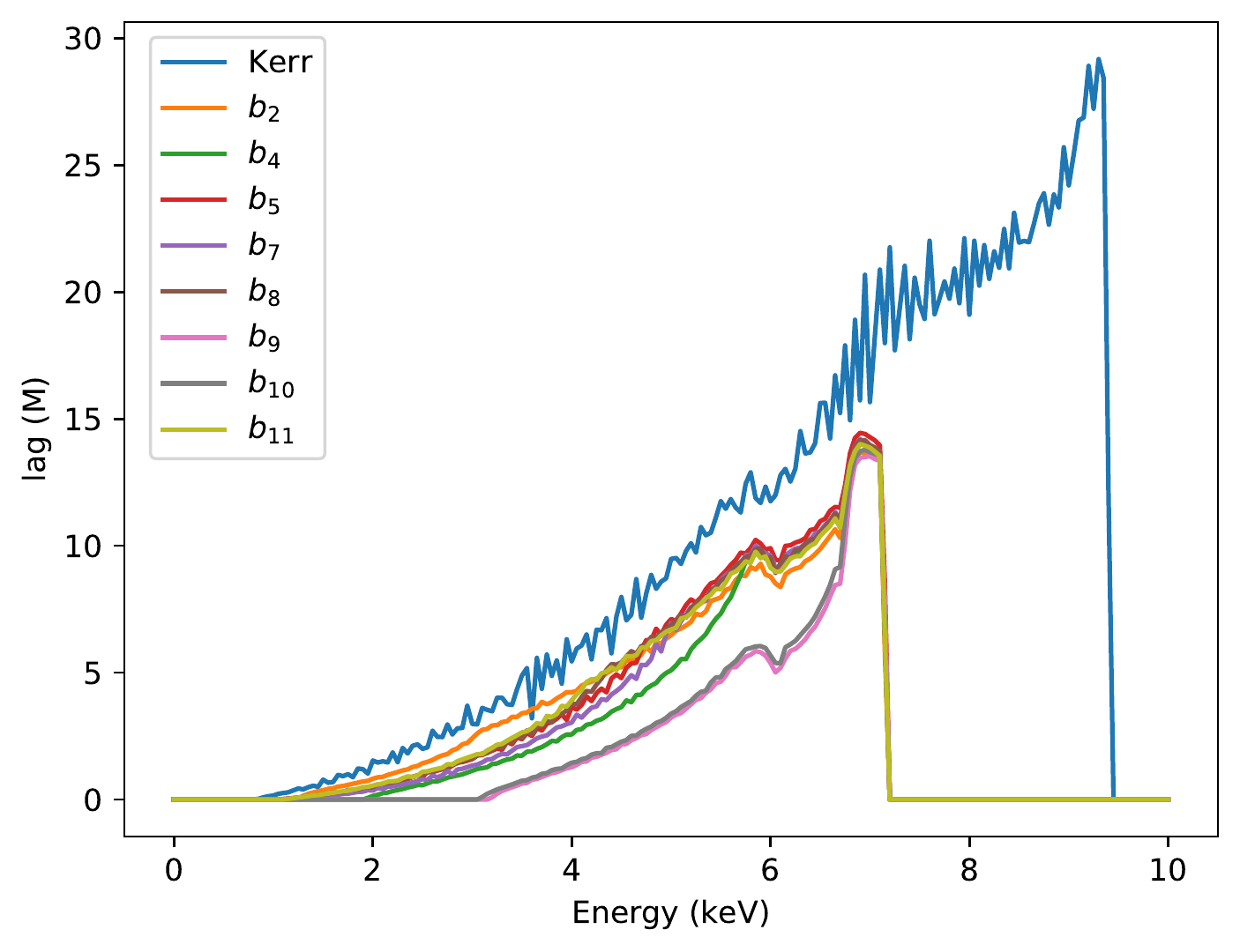}
\end{center}
\vspace{-0.3cm}
\caption{Lag vs energy for different $b_i = 5$. The spin parameter is $0.95$ and viewing angle is $45^{\circ}$. The height of corona is $10$. The frequency range is $(3-4)\times 10^{-3}$ 1/M. \label{energy}}
\end{figure*}

\cb{The energy dependence of the lag is shown in Fig.~\ref{energy}. There is less lag than in the Kerr case for all parameters. The blue part shifts to lower energies if we compare with the Kerr case. The red wings of the parameters $b_2, b_4, b_5, b_7, b_8, b_{11}$ are longer than the parameters $b_9$ and $b_{10}$. Also $b_9$ and $b_{10}$ have a smaller lag than the other ones.\\
For the effect of other parameters such as the BH spin, inclination, height of the corona and so on one can refer to~\cite{Cack}.}


\section{Summary and conclusions \label{summary}}

Iron line reverberation mapping concerns the time lag between coronal photons and photons reflected from the disk. This lag depends on the light travel distances. Thus, this provides us with the opportunity to probe spacetime geometry in strong gravity regimes. 
Here I consider the lamppost coronal geometry; I have an additional height of the corona just above the BH as my parameters. Furthermore, parametrization of the Kerr BH is one way to study deviations from Kerr BH. In this paper I apply reverberation mapping studies to the GB parametrized metric. In GB background I have eight Kerr parameters in addition to the mass and spin parameters of the BH. By the opportunity to study the reverberation associated with the iron line, our GB parameters can be constrained; there is an exception for the parameter $b_{11}$, which introduces degeneracy with the Kerr case. The parameter $b_4$ is also harder to constrain. \\
I already studied this metric by the BH shadow, and the time-integrated iron line.
Time-integrated iron line studies can constrain our Kerr parameters with the exception of $b_{11}$. Notice that our iron line study deals with future observational facilities, but in a reverberation mapping the photon count $10^3$ is for current high quality data. The boundary of the shadow can be slightly altered by the parameters $b_2, b_8, b_9$ and $b_{10}$ but there are no signatures on the shadow shape for the parameters $b_4, b_5, b_7$ and $b_{11}$. 
\cb{The response function, frequency dependence and energy dependence of the lag also discussed in the paper. }
In conclusion, using available current data, the power of the reverberation mapping in the study of strong gravity regimes is clear.

{\bf Acknowledgments} This work is supported by School of Astronomy, IPM, Tehran, Iran.


\end{document}